\documentstyle[twoside,fleqn,espcrc2]{article}

\input epsf
\def\vev#1{\langle#1\rangle}
\def\Dsl{\,\raise.15ex\hbox{/}\mkern-13.5mu D}
\def\cO{{\cal O}}
\def\GeV{\mathop{\rm GeV}\nolimits}
\def\MeV{\mathop{\rm MeV}\nolimits}
\def\mynegskip{\vskip-.32in}

\title{
\vspace*{-15pt}
The $\eta'$ and Cooling with Staggered Fermions}

\author{
G.~Kilcup, J. Grandy and L. Venkataraman
\address{Department of Physics, The Ohio State University,
    174 West 18th Ave,  Columbus, Ohio 43210}
}

\begin{document}

\begin{abstract}
We present a calculation of the mass of the $\eta'$
meson using quenched and dynamical staggered fermions.
We also discuss the effects of ``cooling'', and suggest
its use a quantitative tool.
\end{abstract}

\maketitle

\section{$\eta'$ Lore}

One of the more curious inhabitants of the low energy
spectrum of QCD is the $\eta'$ meson.  Deriving part
of its mass from its kinship with the Goldstone boson
pseudoscalars, the $\eta'$ owes most of its anomalous weight
to its connection to the topological susceptibility.
Specifically, in an $SU(3)$ symmetric
world, one would write
\begin{equation}
m_{\eta'}^2 = m_8^2 + m_0^2
\end{equation}
where $m_8$ is the mass common to all of the octet meson,
and $m_0$ is peculiar to the $\eta'$.  In the chiral
limit, the octet mass $m_8$ vanishes like $\sqrt{m_q}$,
while $m_0$ remains finite.
Ignoring mixing and identifying $m_8^2$ with the average mass-squared
of the physical $\pi$'s, $K$'s and the $\eta$, one derives the
``experimental'' number of $m_0=860$ MeV for $N_f=3$.
To complete the calculation of the ground state hadron
spectrum, one would like to compute this number from a
first principles lattice simulation, free of assumptions
such as the validity of the large $N_c$ expansion.

In the language of quarks, the $\eta'$ is understood
to gain its mass from ``hairpin'' diagrams, where the
quarks annihilate into glue and then reconstitute themselves.
In the language of mesons, this phenomenon can be described by
ascribing a strength $m_0^2$ to this interaction of flavor
singlet mesons.  Iteration of this process leads to a
geometric series which shifts the pole in the propagator
from $m_8^2$ to $m_8^2+m_0^2$.
In the quenched approximation, the series terminates with
the second term.  The quenched $\eta'$ propagator then has
the unphysical form of a sum of a pole and a double pole,
the infrared divergences of which lead to extra quenched
chiral logs.
It is therefore of interest to see if one can distinguish
pole and double pole behavior in a lattice simulation.

\begin{figure}[htbp]
\begin{center}
\leavevmode
\epsfxsize=2.7in
\epsfbox{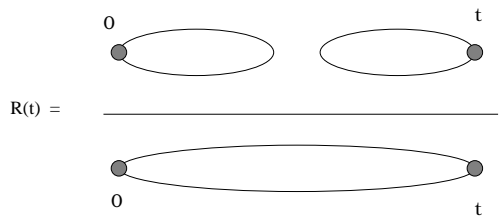}
\end{center}
\caption[]{ Diagrams for the $\eta'$ propagator.}
\label{fig:diag}
\end{figure}
\mynegskip

The quantity of interest is the ratio of the
disconnected and connected diagrams.
\begin{equation}
R(t) = {\langle\eta'(0)\eta'(t)\rangle_{\rm 2-loop}
        \over\langle\eta'(0)\eta'(t)\rangle_{\rm 1-loop}}
\end{equation}
If we compute the ratio using $N_{\rm val}$ valence
quarks and $N_{\rm dyn}$ dynamical quarks, with the
same quark mass for valence and dynamical, then
$R(t)$ should asymptote to
\begin{equation}
R(t) \rightarrow  {N_{\rm val}\over N_{\rm dyn}}
[1 - {Z'\over Z}\exp(-t \Delta m)]
\end{equation}
where $\Delta m = m_\eta'-m_8$, and $Z'$ and $Z$ are the
residues for the creation of the singlet and octet particles,
respectively.
For $N_{\rm dyn}=0$, i.e. in the quenched approximation,
$R(t)$ should never asymptote, but instead rises linearly
with slope $2 m_0^2/m_8$.

This approach has been taken with quenched configurations
and valence Wilson fermions in ref. \cite{EtaWilson}.
Here we extend the analysis to staggered fermions, and
use quenched and dynamical configurations.

\section{Implementation}

This calculation used three ensembles of lattices of size
$16^3\times32$, one quenched and two including dynamical quarks.
The dynamical ensembles were ``borrowed'' from the Columbia group,
and have been used previously in our $B_K$ analysis \cite{DynBK},
and the calculation of $f_B$ by the MILC collaboration \cite{MILCfB}.
These were generated using molecular dynamics with $N_f=2$ staggered
fermions with quark masses $m_{\rm dyn}=.01$ and $.025$.
The quenched lattices were generated on the Cray-T3D at the
Ohio Supercomputer Center (OSC), using 3-subgroup $SU(2)$
updates, with 4 to 1 mixture of over-relaxed and  heatbath.
We saved configurations with 2000 sweeps, and obtained an
average update time of 5 microseconds per link on 16 processors.

\vskip -.2in

\begin{table}[hbt]
\caption{\ \ \ \ \ \ \ The Statistical Ensemble}
\label{tab:ensemble}
{\centerline{
\begin{tabular}{cccc} \hline
$\beta$:& 6.0 & 5.7 & 5.7 \\
$N_f$:& 0 & 2 & 2 \\
$m_{dyn}$:& $\infty$ & .025 & .01 \\
$N_{\rm samp}$: & 32 & 35 & 49 \cr
\hline
\end{tabular}
}}
\end{table}
\vskip -.2in

We computed staggered propagators on the 32 node
T3D at OSC and on the 256 node T3D at Los Alamos's Advanced
Computer Laboratory.  We ran our fixed size problem on sets
of processors ranging from 16 nodes to 128 nodes, and found
the per node performance to be essentially independent of the
partition size over this range.  By hand coding the inner loops
in CAM, the Cray T3D assembly language, we obtained a sustained
performance of 45 Mflops per node, including I/O time.

To create and destroy the $\eta'$ we used the staggered
flavor singlet operator $\bar Q(\gamma_5\otimes I)Q$.
This is a distance 4 operator, which we made gauge invariant
by putting in explicit links, and averaging over the 24 paths
across the edges of the hypercube.

To obtain the one-loop contractions of this operator
(the denominator of fig. \ref{fig:diag}), we computed
two types of propagators.  One type used a noisy source
which was nonzero on two timeslices $t=0$ and $t=1$.
This source was a random phase $\eta_x=e^{i\theta(x)}$
for each color and each site, such that in the average
over noise samples, $\vev{\eta_x\eta^{\dag}_y} = \delta_{xy}$.
The second type of propagator used a source obtained by
transporting the noise $\eta_x$ across the diagonal of
the hypercube and putting the phase appropriate for the
$\eta'$ operator.  Thus the operator which created the
$\eta'$ was the spatial sum over a double timeslice of
the hypercube operator, plus noise terms which vanish
on average.  We took two noise samples per configuration.

For the two-loop contractions we used source $\eta$ with
$U(1)$ noise at every site on the lattice.  Then solving
$(\Dsl + m)\phi = \eta$ we estimate the propagator as
$G_{xy} = \vev{m\phi_x\phi^{\dag}_y}$.  We note that this
estimator is orders of magnitude better than the
alternative $G_{xy} = \vev{\phi_x\eta^{\dag}_y}$,
but it only available when the sites $x$ and $y$
are separated by an {\em even} distance.  For this
reason we restricted our attention to the pseudoscalar
operator, and neglected the axial vector (which is
distance 3).  For each color we used 16 noise samples
on a doubled lattice, or an effective number of 96
noise vectors per mass.

\section{Results}

\begin{figure}[htbp]
\begin{center}
\leavevmode
\epsfxsize=2.7in
\epsfbox[0 150 750 400]{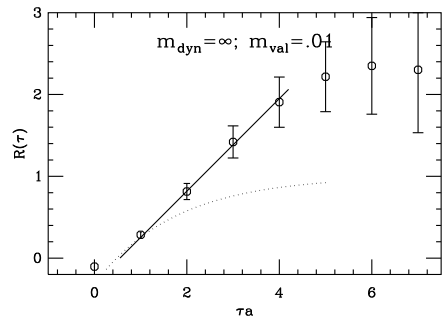}
\end{center}
\caption[]{Quenched $R(\tau)$.}
\label{fig:r_quenched}
\end{figure}
\mynegskip

\begin{figure}[htbp]
\begin{center}
\leavevmode
\epsfxsize=2.7in
\epsfbox[0 150 750 400]{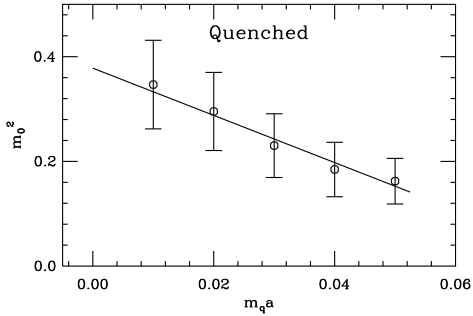}
\end{center}
\caption[]{Quenched $m_0^2$. }
\label{fig:m0_q}
\end{figure}

Figure \ref{fig:r_quenched} shows the measured ratio $R(\tau)$
in the quenched ensemble at a valence quark mass $m_q=.01$.
In principle, the sickness of the quenched approximation
should manifest itself in an unending linear trend in the data.
By contrast, the dotted curve is the exponential one would expect
if there were $N_f=4$ active dynamical flavors.
Clearly the data rule out $N_f=4$, and are
consistent with the quenched form, but from the data
alone one couldn't rule out a nonpathological behavior
with $N_f=2$ or fewer flavors.
Extracting the slope from such curves for several
quark masses, we obtain the values for $m_0$ plotted
in figure \ref{fig:m0_q}

\mynegskip

\noindent
Extrapolating linearly to $m_q=0$ and rescaling to the
most relevant case of $N_f=3$ degenerate flavors, we obtain
\begin{equation}
m_0^2(N_f=3) = (1050\pm170 \MeV)^2 ({a^{-1}\over 2 \GeV})^2,
\end{equation}
which is compatible with the ``experimental'' number
quoted above.

\begin{figure}[htbp]
\begin{center}
\leavevmode
\epsfxsize=2.7in
\epsfbox[0 150 750 400]{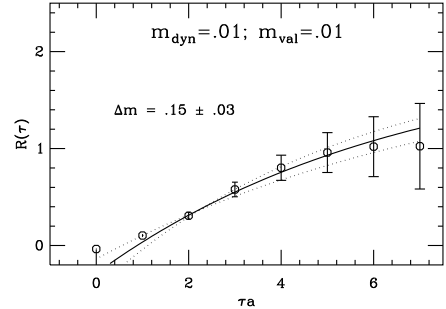}
\end{center}
\caption[]{Dynamical $R(\tau)$. }
\label{fig:r010_d}
\end{figure}

In the presence of $N_f=2$ dynamical fermions, we expect
$R(\tau)$ to asymptote to the constant 2.
Figure \ref{fig:r010_d} shows the result for $m_{dyn}=m_{val}=.01$.
To fit it to the exponential form may be putting more weight
on the data than it should bear, but its behavior does
stand in clear contrast to the quenched data.
If we ignore our theoretical prejudice and simply press ahead
to repeat the linearized analysis as in the quenched case,
we find the result
\begin{equation}
m_0^2(N_f=3) = (780\pm50 \MeV)^2 ({a^{-1}\over 2 \GeV})^2,
\end{equation}
On the other hand, if we do the correct job and fit
to the exponential form, we can extract $m_{\eta'}a$
at two points, $m_{dyn}=m_{val}=.01$ and $m_{dyn}=m_{val}=.025$,
finding $.488\pm.030$ and $.676\pm.040$ for the bare lattice
numbers respectively. Extrapolating to $m_q=0$ and rescaling
to $N_f=3$ we end up with
\begin{equation}
m_0^2(N_f=3) = (730\pm250 \MeV)^2 ({a^{-1}\over 2 \GeV})^2
\end{equation}

\mynegskip

\section{Cooling}

\begin{figure}[htbp]
\begin{center}
\leavevmode
\epsfxsize=2.7in
\epsfbox[0 150 750 400]{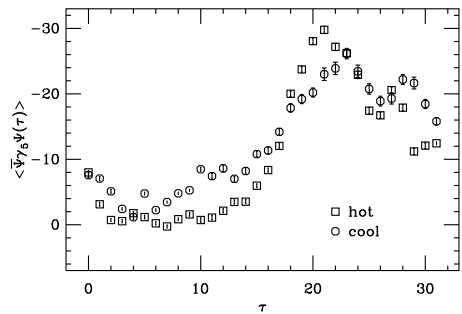}
\end{center}
\caption[]{Effect of 50 cooling steps.}
\label{fig:qhotcool}
\end{figure}

In a closely related study, we have ``cooled'' the dynamical
($m_q=.01$) to count large instantons.
Dividing the configurations into bins according to their
cooled topological charge, we compute the $\eta'$ mass
on two subensembles, one with $n=0$,1 or 2 instantons,
and one with $n=3$,4 or 5 instantons.
Like the authors of ref. \cite{WilsonCool} we find that
$m_0^2$ mass is larger (by a factor 2) in the subensemble
with larger topological charge, in accord with the idea that
the $\eta'$ gets its mass from instantons.
This result can be understood by looking at the
pseudoscalar expectation value in particular configurations.
Figure \ref{fig:qhotcool}
shows what happens to the $\eta'$ operator as one cools.
The correlation evident in the plot persists as a function
of cooling time, even out to the point where all that
remains are smooth instantons.  We conclude that even in
the hot configuration, the large features in the $\eta'$
are due to underlying instantons.
\mynegskip

This observation invites one to explore a method which
could speed up this and other calculations.
Any observable we care to compute can be written as
\begin{equation}
\label{eqn:cool}
\vev{\cO} =
\vev{\cO}_{cool} \times \big({\vev{\cO} \over \vev{\cO}_{cool}}\big),
\end{equation}
where $\cO_{cool}$ indicates observables computed on cooled
configurations.  In general $\cO_{cool}$ are {\em much}
cheaper to compute, and have smaller statistical fluctuations.
If the hot and cool observables are strongly correlated,
it may then be the case that for a fixed amount of computing
one can obtain the RHS of eqn. \ref{eqn:cool} to better
precision than the LHS.  The method amounts to a type of
renormalization group transformation, where $\cO_{cool}$
is supposed to retain the long distance physics.

\begin{figure}[htbp]
\begin{center}
\leavevmode
\epsfxsize=2.7in
\epsfbox[0 150 750 400]{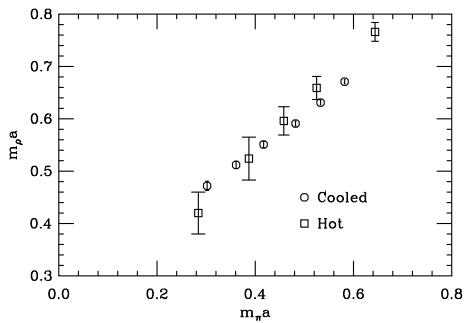}
\end{center}
\caption[]{$\rho$ and $\pi$ masses.}
\label{fig:rhopi}
\end{figure}

In a pilot study we tried cooling by only two steps.
Figure \ref{fig:rhopi} is a scatterplot of $m_\pi$ and
$m_\rho$ for a variety of quark masses in both the hot
and cold ensembles.  Extrapolating to zero quark mass,
the hot and cool ensembles give compatible rho masses,
but evidently the cool data are much more precise.
Further, if we compare at fixed pion mass, the cool data
are a factor of 3 to 4 cheaper in terms of CG iterations.
\mynegskip

\begin{figure}[htbp]
\begin{center}
\leavevmode
\epsfxsize=2.7in
\epsfbox[0 150 750 400]{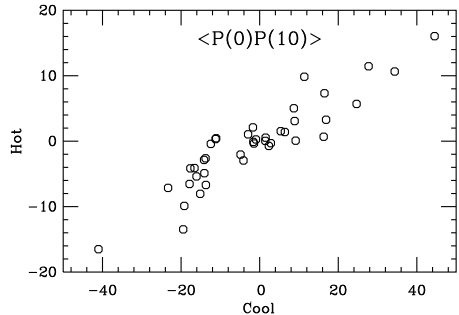}
\end{center}
\caption[]{Hot and cool observables.}
\label{fig:corr}
\end{figure}

Picking quark masses which are close to each other
in figure \ref{fig:rhopi} ($m_q=.01$ hot, and $m_q=.04$ cool),
we can look for the correlations in more complicated observables,
such as the disconnected $\vev{\eta'(0)\eta'(t)}$ two-point function.
Figure \ref{fig:corr} shows a scatterplot of the data for $t=10$.
As the dispersions show, both variables are statistically
consistent with zero, i.e. we are well into the noise.
Still, the two observables are strongly correlated,
and we can determine their ratio more precisely than we can
determine either variable alone.

\mynegskip

\section{Conclusions and Future Directions}

We have shown that the $\eta'$ mass can be computed
using staggered fermions
While the data are too imprecise for definitive
conclusions, it is at least plausible that the
quenched $\eta'$ propagator is pathological,
while the dynamical one is well behaved.
To compute the study we want to repeat the calculation
with smeared sources to reduce the  effect of higher excitations.
We have also suggested a style of calculation which
may be of some use when computing expensive,
propagator-intensive observables.  Our initial choice
of two large cooling steps is unlikely to be optimal;
to fully test the idea one should try better
smoothing algorithms, e.g. using an improved action.

\leftline{\bf Acknowledgment}
These calculations were performed on the Cray-T3D's
at the Ohio Supercomputer Center and at the Advanced
Computer Laboratory.  We thank OSC and ACL for early
access to these machines.

\end{document}